\documentclass[prb,twocolumn,showpacs,superscriptaddress,floatfix,amsfonts,amsmath]{revtex4}

\usepackage{graphicx,subfigure}

\newcommand{\be}{\begin{equation}}
\newcommand{\ee}{\end{equation}}
\newcommand{\bea}{\begin{eqnarray}}
\newcommand{\eea}{\end{eqnarray}}
\newcommand{\down}{\downarrow}
\newcommand{\dn}{\downarrow}
\newcommand{\up}{\uparrow}
\newcommand{\f}{\frac}
\newcommand{\cobaltchloride}{\rm{CoCl}_2\cdot 2\rm{H}_2\rm{O}}
\newcommand{\cobaltniobate}{\rm{CoNb}_2\rm{O}_6}

\begin{document}

\title{Spectral signatures of magnetic Bloch oscillations in one-dimensional easy-axis ferromagnets}
\author{Sergey \surname{Shinkevich}}
\author{Olav  F. \surname{Sylju{\aa}sen}}
\affiliation{Department of Physics, University of Oslo, P.~O.~Box 1048 Blindern, N-0316 Oslo, Norway}

\date{\today}

\begin{abstract}
Domain-walls in a one-dimensional gapped easy-axis ferromagnet can exhibit Bloch oscillations in an applied magnetic field. We investigate how exchange couplings modify this behavior within an approximation based on noninteracting domain-wall bound states.  In particular, we obtain analytical results for the spectrum and the dynamic structure factor, and show where in momentum space to expect equidistant energy levels, the Wannier-Zeeman ladder, which is the spectral signature of magnetic Bloch oscillations.  We compare our results to previous calculations employing a single domain-wall approximation, and make predictions relevant for the material $\cobaltchloride$.
\end{abstract}

\maketitle

\section{introduction}
Quantum mechanics predicts that a particle in a periodic potential will undergo oscillatory motion in response to a {\em constant} force. This rather counterintuitive phenomenon, known as Bloch oscillations (BO), was controversial for a long time, but has now been experimentally demonstrated in very clean semiconductor superlattices\cite{BOsemiconductors} and Bose-Einstein condensates\cite{BOcondensates}.

Can Bloch oscillations also exist in magnetic systems? Kyriakidis and Loss\cite{KyriakidisLoss} discussed this possibility. They considered a system where the particle is a propagating domain-wall excitation in an easy-axis one-dimensional ferromagnet, and concluded that magnetic BO should indeed exist. In particular, the blue crystalline material $\cobaltchloride$ was identified as a promising candidate for observing magnetic BO. Searches using neutron scattering were performed\cite{Christensen,Montfrooij}, but did not find evidence of BO.

Another very similar system where one can expect BO of magnetic domain-walls is the Ising model in a magnetic field having both longitudinal and transversal components\cite{Cai}. Such a model is believed to be realized in $\cobaltniobate$ where indeed an intriguing frequency spectrum has recently been observed\cite{Coldea}.  However, in the region of momentum space where one expects to find the quantum mechanical spectral signature of BO --- a spectrum with equidistant energy levels, the so-called Wannier-Zeeman ladder (WZL) --- the spectral weight in the experiment in Ref.~\onlinecite{Coldea} is dominated by a strong feature attributed to additional couplings in the Hamiltonian, a ``kinetic bound state'', stabilized by next-nearest neighbor interactions\cite{Coldea,Kjall}. Thus it appears that additional terms in the Hamiltonian prevent BO in $\cobaltniobate$.

 This might also be the case in $\cobaltchloride$ where the ``kinetic bound state'' will be generated by a nearest neighbor spin flip exchange coupling that, indeed, is present in $\cobaltchloride$\cite{Montfrooij}, but neglected in Ref.~\onlinecite{KyriakidisLoss}. It is the main goal of this article to investigate its influence on the WZL in $\cobaltchloride$.




Our results show that the WZL is present in certain regions of momentum space also in the presence of the exchange couplings. However, the neutron scattering spectral weight of the WZL in $\cobaltchloride$ is less than 1\% of the total spectral weight at these momenta, thus making it difficult to observe the WZL at zero temperature in inelastic neutron scattering experiments.

At finite temperatures, the neutron scattering signatures of the WZL are more pronounced. We find that a relatively high temperature is favorable as the number of domain-walls performing BO are exponentially suppressed with temperature below the largest ferromagnetic coupling.
Unfortunately, a high temperature leads also to collisions of domain-walls that destroy BO. For $\cobaltchloride$, we find that this collision rate is determined by the velocity of the ``kinetic bound state'' and the distance between domain-walls bound states that gets smaller as the temperature increases. A way to alleviate this is to increase the applied magnetic field that leads to a reduced collision rate. However, a large magnetic field makes the intrinsic signature of BO weaker as it reduces the BO amplitude. In searching for a compromise, we find significant neutron scattering signatures of BO at finite frequencies in $\cobaltchloride$ at the temperature  $T=J_z/2$, magnetic field $h_z=0.2J_z$, and momentum transfer $q=\pi/2$ where the first Bloch mode at $\omega_B=2h_z$ carries about 12\% of the total spectral weight.

While we focus on the material parameters for $\cobaltchloride$, our results are analytic and can, with minor efforts, also be used for aiding searches for magnetic BO in other similar materials.

\section{Hamiltonian}
We start with the spin-1/2 \emph{XYZ} ferromagnetic Hamiltonian for a chain with nearest-neighbor coupling in a magnetic field
\be
    H = - \sum_{i} \left( J_x S^x_i S^x_{i+1} + J_y S^y_i S^y_{i+1} + J_z S^z_i S^z_{i+1} +h_z S^z_i \right),
\ee
which can be written as
\be
    H = H_z + H_a + H_\perp,
\ee
where
\bea
    H_z & = & -J_z \sum_{i} S^z_i S^z_{i+1}  - h_z \sum_{i} S^z_i, \nonumber \\
    H_a & = & -J_a \sum_{i} \left( S^+_i S^+_{i+1} + S^-_i S^-_{i+1} \right), \nonumber \\
    H_{\perp} & = & - J_\perp \sum_{i} \left( S^+_i S^-_{i+1} + S^-_i S^+_{i+1} \right) \nonumber
\eea
with $J_a=(J_x-J_y)/4$ and $J_{\perp}=(J_x+J_y)/4$, and $S^{\pm}_i= S^x_i \pm i S^y_i$ are the usual raising and lowering operators.
We will assume that $J_z$ is by far the largest coupling thus the system has an easy-axis and its behavior is Ising-like. The ferromagnetic coupling causes neighboring spins to align their \emph{z}-components, and with a large $J_z$, the excitation energy of a state is mainly dependent on the number of anti-aligned spin neighbors, or domain-walls. Each domain wall costs an energy of $J_z/2$ (our $J_z$ is twice that of Ref.~\onlinecite{TorranceTinkham}). Thus the ground state is approximately the ferromagnetic state where all spins are aligned along the \emph{z}-axis. In the absence of a magnetic field and other couplings, the first excited state has a single domain-wall, see Fig.~\ref{Fig_1domain}. This domain-wall can be placed between any of the spins, implying a macroscopic degeneracy. The $H_a$-term lifts this degeneracy resulting in a band dispersion describing the dynamics of a single domain-wall. This mode was first predicted by Villain\cite{Villain} and was subsequently observed in neutron scattering experiments\cite{Villainmodeobs1,Villainmodeobs2}.

Kyriakidis and Loss treated a single domain-wall in the presence of a finite magnetic field and predicted BO\cite{KyriakidisLoss}. The BO are caused by $H_a$ together with the magnetic field that causes the domain wall to oscillate. While the single domain-wall approximation is presumably good on short time-scales where collisions between domain-walls can be ignored, it will break down at longer time scales. In a finite magnetic field, this time-scale is likely to be very short as a domain-wall and an anti-domain-wall are closely bound together. The energy cost of a domain of spins anti-aligned with the field is proportional to the magnetic field times the number of spins in the domain. Thus the magnetic field induces a linear potential between a domain-wall and an anti-domain-wall which confines them in a bound state. Therefore the low energy excitations will not be isolated single domain-walls, but rather bound states of domain-wall/antidomain-wall pairs that define the boundaries of a spin cluster of overturned spins, see Fig.~\ref{Fig_2domain}. The far-infrared light absorption experiments on the quasi-one-dimensional material $\cobaltchloride$ in a magnetic field have been explained in terms of such spin cluster excitations\cite{TorranceTinkham,Fogedby}. In the bound state picture, $H_a$ and the magnetic field cause the bound state to shrink and expand. This is what gives rise to the BO and the WZL.

Going beyond the single domain-wall approximation also allows the inclusion of the spin-flip exchange Hamiltonian $H_{\perp}$. The action of $H_\perp$ on the single domain-wall state shown in Fig.~\ref{Fig_1domain} produces a high energy state having three domain-walls. In contrast when $H_\perp$ acts on the minimal bound state --- a single overturned spin, see Fig.~\ref{Fig_2domain} right --- it can move the whole bound state without introducing extra domain-walls, thus $H_\perp$ acts directly in the low energy Hilbert space of a single bound state. This minimal bound state is the analogue to the ``kinetic bound state'' in $\cobaltniobate$.

In zero magnetic field, the predictions for the neutron scattering dynamic structure factor were shown to be independent\cite{Nagler} of whether one considered noninteracting domain-walls\cite{Villain} or bound states\cite{IshimuraShiba}. However, as argued above, this does not hold in a finite magnetic field, at least not on longer time scales. In this article we will treat the single bound state exactly in the low energy subspace to two domain-walls and assume that thermally the system can be well approximated by a noninteracting gas of such bound states.
\begin{figure}
\begin{tabular}{c}
 $\up \up \up \up \up \up \dn \dn \dn \dn \dn \dn $
\end{tabular}
\caption{A spin state with a single domain-wall.}
\label{Fig_1domain}
\end{figure}

\begin{figure}
\begin{tabular}{c}
 $\up \up \dn \dn \up \up \up \up \dn \up \up \up $
\end{tabular}
\caption{Two bound states.}
\label{Fig_2domain}
\end{figure}


\section{quantum mechanics of a single bound state}
\noindent
Let us represent a bound state of a domain-wall and an anti-domain-wall as the state
\be
|j, l \rangle =| \ldots \up \up \underbrace{\mathop{\down}^j \! \down \ldots \down}_l \up \up \ldots \rangle,
\ee
where the index $j=1,2,\ldots N$ gives the starting position of the down-spin cluster and $l=1,\ldots N$ describes its length. $N$ is the total number of spins in the chain, which we will take to be a macroscopic number. We will extend this representation of states to $l=0$ in order to also include the ferromagnetic state $|j,0 \rangle$, which is independent of $j$.
The action of the Hamiltonian on such a state can be written as
\bea
    &H |j,l \rangle &= (1-\delta_{l,0}) \Bigl \{ J_z |j,l \rangle + h_z l |j,l \rangle \nonumber \\
      & &- J_a \bigl[ |j,l+2 \rangle + |j-2,l+2 \rangle \nonumber \\
      & &+ \left(|j,l-2 \rangle + |j+2,l-2 \rangle\right)(1-\delta_{l,2})(1-\delta_{l,1}) \bigr] \Bigr \} \nonumber \\
      & &- J_a (|j,2 \rangle \delta_{l,0} + |j,0 \rangle \delta_{l,2}) \nonumber \\
      & &- J_\perp (|j+1,1 \rangle + |j-1,1 \rangle) \delta_{l,1},
\eea
where we have neglected terms that result in more than two domain-walls.

We consider a system with periodic boundary conditions. This ensures translational invariance and the total momentum of the bound state will be a good quantum number. It is thus convenient to express the Hamiltonian in the momentum basis $|p,l \rangle =e^{-ipl/2} \sum_j e^{-i pj}|j,l \rangle$, where $p$ denotes the total momentum of the bound state (we use units where the lattice spacing is one). Note that the ferromagnetic state necessarily has zero momentum $|p=0,0\rangle$. In the momentum basis, the Hamiltonian is diagonal in $p$ and acts as follows
\begin{align}
    H |p,l\rangle &= (1-\delta_{l,0}) \bigl \{ [J_z + h_z l -2J_\perp \cos{p} \, \delta_{l,1} ] |p,l\rangle \nonumber \\
    &- 2J_a \cos{p}\,[ |p,l+2\rangle + (1-\delta_{l,1})(1-\delta_{l,2}) |p,l-2 \rangle]\bigr \} \nonumber \\
    &- J_a ( |p,2 \rangle \delta_{l,0}+ |p,0 \rangle \delta_{l,2}) \delta_{p,0}.
\end{align}
Because $H_a$ flips two spins, the state sectors with even and odd values of $l$ are decoupled.  Note that $H_{\perp}$ only affects the odd $l$ sector. This follows from the fact that $H_{\perp}$ is only non-zero when it acts on the state with a single down-spin, the action on all other states produces more domain-walls, see Fig.~\ref{Fig_H_action}.
\begin{figure}
\begin{tabular}{lllrp{0.3cm}|p{0.3cm}lllr}
                           & $H_a$         &                               &       & & &                            & $H_{\perp}$    &                               &       \\
$\up \up \up \up \up \up $ & $\rightarrow$ & $\up \up \dn \dn \up \up$ & $,+2$ & & & $\up \up \up \up \up \up $ & $\rightarrow$ & 0 & \\
$\up \dn \dn \up \up \up $ & $\rightarrow$ & $\up \dn \dn \dn \dn \up$ & $, 0$ & & & $\up \dn \dn \up \up \up $ & $\rightarrow$ & $\up \dn \up \dn \up \up$ & $,+2$ \\
$\up \up \dn \up \up \up $ & $\rightarrow$ & $\up \up \dn \dn \dn \up$ & $, 0$ & & & $\up \up \dn \up \up \up $ & $\rightarrow$ & $\up \up \up \dn \up \up$ & $, 0$
\end{tabular}
\caption{Examples of the actions of $H_a$ on particular states from the 0 and 2 domain-wall sectors. Only cases that yield the minimal amount of increase in domain walls are shown. Operating with $H_a$ are shown on the left, while the right hand side shows the effects of operating with $H_{\perp}$. The increase in the number of domain walls is indicated to the right of each process. Note that $H_a$ has the ability to move domain walls without increasing their number when acting on a state with one or more domain walls (left side, the two lowest processes), while $H_\perp$ lacks this ability with the exception that it can move a single overturned spin without creating new domain walls (right side, the lowest process).  }
\label{Fig_H_action}
\end{figure}

In order to solve the eigenvalue problem, we first parametrize the energy eigenvalues as
\be
  E_{n}(p) = J_z +  h_z \mu_n,
\ee
where $p$ is the total momentum, $n$ is a positive integer, which labels the internal excitation mode of the bound state, and $\mu_n$ depends on $p$ and coupling constants of the Hamiltonian. Because the excitation spectrum separates into distinct sectors, we will reserve the odd(even) $n$ values for labeling the energy levels in the $l$ odd(even) sector.
The energy of the $n$'th mode is found by determining the dimensionless quantity $\mu_{n}$ from the following equations:
\begin{eqnarray}
  \f{J_{-(\mu_n+1)/2} (x)}{J_{1-(\mu_n+1)/2}(x)} & = & z, \quad {\rm n \in odd,}
  \label{eq_nu_odd} \\
  \f{J_{-{\mu_n}/2} (x)}{J_{1-\mu_n/2} (x)} &=& 0,  \quad {\rm n \in even,}
  \label{eq_nu_even}
\end{eqnarray}
with $J_\nu(x)$ being the Bessel function of the first kind of order $\nu$. The upper(lower) equation is obtained by considering the odd(even) $l$ sector. Here, we have introduced the new variables
\be
  x = \frac{2 J_a \cos{p}}{h_z}, \quad  z = \frac{J_\perp}{J_a}.
\ee
The even sector equation (\ref{eq_nu_even}) is only valid for $p \neq 0$. The $p=0$ case will be considered separately below.

\begin{figure}[tbp]
\begin{center}
  \includegraphics[clip,width=7cm]{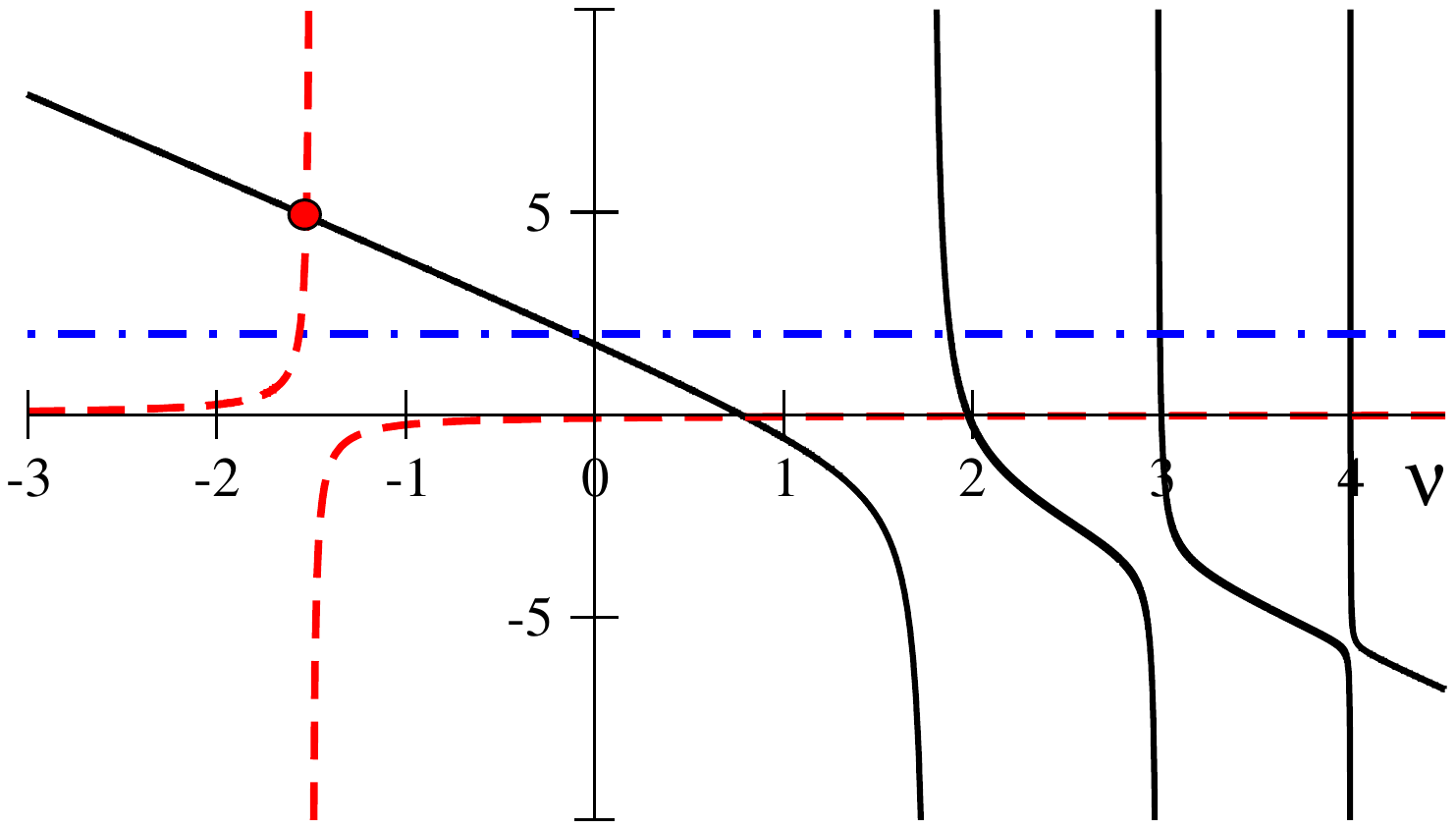} 
  \caption{(Color online) The behavior of the ratio of Bessel functions $J_{-\nu}(x=1)/J_{1-\nu}(x=1)$ as a function of $\nu$ is shown as the black solid curve. The blue dot-dashed line marks the value $\gamma=2$, while the red dashed curve shows the right hand side of Eq.~(\ref{pzeroroots}) for $y=3$. The red circle indicates the crossing point that gives the lowest energy solution $\nu_0=\mu_0/2$ of the $p=0$ even sector.}
  \label{generalformplot}
\end{center}
\end{figure}

\begin{figure}[tbp]
\begin{center}
    \includegraphics[clip,width=7cm]{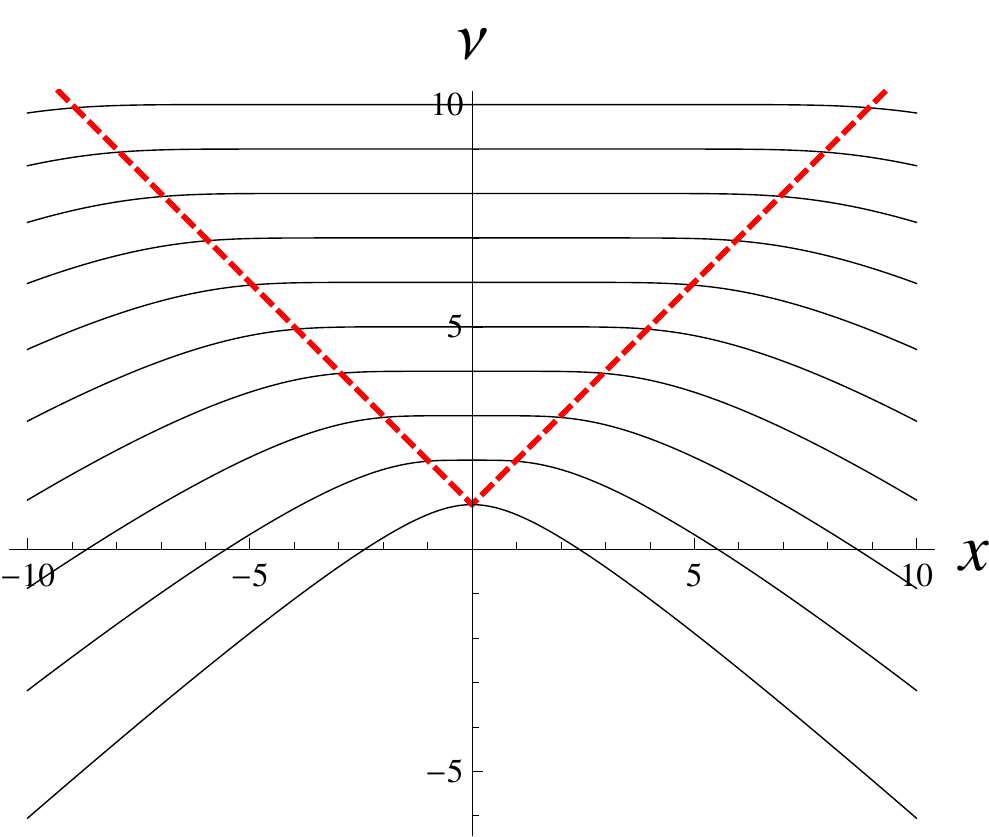} 
   \caption{(Color online) Solutions $\nu$ of Eq.~(\ref{generalform}) as a function of $x$ for $\gamma=0$. The red dashed line is the line $\nu=1+|x|$.}
   \label{rootseven}
  \end{center}
\end{figure}

Both equations are of the form
\be
    \f{J_{-\nu}(x)}{J_{1-\nu}(x)} = \gamma, \label{generalform}
\ee
where $\gamma$ is a constant. Analytical solutions of this equation has been found in some limits\cite{Rutkevich}.
To get an intuitive picture of the solutions of Eq.~(\ref{generalform}), we have plotted the left hand side of the equation for a particular value of $x$ in Fig.~\ref{generalformplot} as the black solid curve. For $\gamma \to 0$, relevant for the even sector and the small $z$ limit of the odd sector, the solutions are gotten by the zero crossings of the curve. We see that they occur almost exactly at positive integer values of $\nu$, except for the lowest value which is somewhat below 1.
In Fig.~\ref{rootseven}, we show these values of $\nu$ as a function of $x$ for $\gamma=0$.
To a very good approximation, $\nu$ is a positive integer as long as $\nu \geq 1+|x|$.
For the even $l$ sector where $\mu=2\nu$, this implies that the solutions $\mu_n$, $n \in \{2,4,6,\ldots\}$ are to a good approximation {\em even} integers $\mu_n=n$ when $n \geq 2|x|+2$. For lower values of $n$, $\mu_n$ is generally lower and depends on $x$.
\begin{figure}[tbp]
\begin{center}
    \includegraphics[clip,width=7cm]{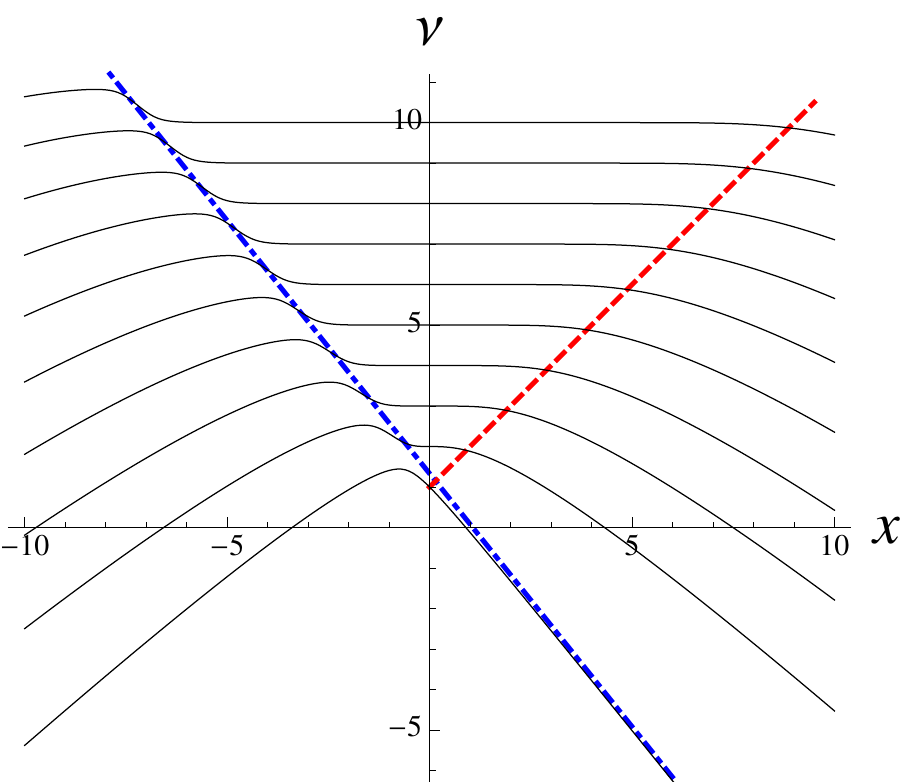} 
   \caption{(Color online) Solutions $\nu$ of Eq.~(\ref{generalform}) as a function of $x$ for $\gamma=2$. The dot-dashed blue line has slope $(\gamma+\gamma^{-1})/2$ and the red dashed line is the line $\nu=1+x$ for positive $x$.}
   \label{rootsodd}
  \end{center}
\end{figure}
For small $\gamma$, relevant for the odd $l$ sector where $\mu = 2\nu-1$ when $z$ is small, the solutions are very similar to the case $\gamma=0$. This means that the solutions $\mu_n$, $n \in \{1,3,5,\ldots\}$ are odd integers $\mu_n=n$ for $n \geq 2|x|+1$. The qualitative effects on the solutions $\nu$ of changing $\gamma$ for a fixed value of $x$ can be inferred from Fig.~\ref{generalformplot}.  While the higher energy levels do not change substantially for this value of $x$, the lowest energy level decreases with increasing $\gamma$. In Fig.~\ref{rootsodd}, we have plotted the solutions of Eq.~(\ref{generalform}) as functions of $x$ for a fixed value of $\gamma=2$. For positive $x$, the lowest energy state decreases with increasing $x$.  This decrease becomes linear at large $x$ and its slope can be found by writing the Bessel function ratio $J_{-\nu}(x)/J_{1-\nu}(x)=-[\nu/x+\f{d}{dx} \ln(J_{-\nu}(x))]^{-1}$ and using the asymptotic expression\cite{Abramowitz}
\be
J_{-\nu}(-\nu \theta) \sim \f{\theta^{-\nu} e^{-\nu \sqrt{1-\theta^2}}}{\sqrt{-2\pi \nu} (1-\theta^2)^{1/4} (1+\sqrt{1-\theta^2})^{-\nu}}
\ee
valid for $\nu \to \infty$ and $0<\theta <1$. We find that the lowest energy state solution decreases as $\nu \sim -(\gamma+\gamma^{-1})x/2$ for $x \to \infty$ and $\gamma > 1$. A line with this slope is overlaid on the plot in Fig.~\ref{rootsodd} for $\gamma=2$. The large $x$ limit corresponds to the limit of vanishing magnetic field. Using the asymptotic result and the definitions of $\gamma$ and $x$, we get for the lowest excited state in the limit of vanishing $h_z$, $E = J_z +h_z (2\nu-1) \sim J_z - 2J_{\perp}(1+(J_a/J_{\perp})^2)\cos p$ corresponding to a spin wave excitation. This result can also be obtained by second order perturbation theory in the limit of vanishing magnetic field when $J_z \gg J_\perp \gg J_a$.

Extrapolating the spin-wave line to negative $x$, we see that it coincides smooth step-like behaviors of the energy levels.  These steps become sharper and higher as $\gamma$ is increased and become a step discontinuity of unit one for $\gamma \to \infty$.
This feature restricts the existence of the WZL for odd $n$ to $n > (\gamma+\gamma^{-1})|x|+1$ for $x<0$ and $\gamma>1$.

From these considerations, it follows that a momentum-independent WZL
\begin{equation}
    E_{n} = J_z + h_z n,
\end{equation}
with integer $n$, is restricted to high energies where
\be
       n > 1+ \f{2J_a}{h_z} \times
        \begin{cases} 2,  & J_\perp \leq J_a, \\
			        J_\perp/J_a+J_a/J_\perp, & J_\perp >J_a. \end{cases}
\ee
For $p=0$, the sector with domain walls of even $l$ will also couple to the ferromagnetic state. Thus,
at $p=0$, the equation for the energy levels in the even sector is different from the one at $p \neq 0$, Eq.~(\ref{eq_nu_even}). For $p=0$, we get
\be
 \f{J_{-\mu_{n}/2} (x_0) }{J_{1-\mu_{n}/2}(x_0)}= - \f{x_0}{4(y+\mu_n)},  \label{pzeroroots}
 \ee
where $y=J_z/h_z$ and $x_0= 2 J_a/h_z$. A similar equation including also the effects of an optical phonon at $p=0$ was obtained in Ref.~\onlinecite{Fogedby}.
The equation for $p=0$ has an additional negative solution $\mu_0 <0$ well separated from the other positive solutions. This solution arises from the singular behavior of the left hand side in the vicinity of $\mu = -y$. Figure~\ref{generalformplot} shows a graphical solution of Eq.~(\ref{pzeroroots}) with $\mu=2\nu$ for fixed values $x_0=1$ and $y=3$. Because of the rapid variations of the ratio of Bessel functions on the left hand side around positive even integer values of $\mu$, we see that a relatively small finite value of the right hand side changes only slightly the even integer solutions found for $p \neq 0$ in the even sector.
However, at low energies, there is a crucial difference. The right hand side has a singularity at $\mu=2\nu =-y$. The left hand side is positive for negative $\mu$ and increases as $\mu \to -\infty$ while the right hand side is negative for $\mu > y$ before it changes sign as $\mu$ passes $y$. Thus somewhere below this singularity a negative solution $\mu_0$ will occur. This means that the ground state energy will be negative as $J+h_z \mu=0$ for $\mu=-y$. For $J_a \ll J_z$ we find for the lowest energy solution approximately
\bea
   E_0 &=&  J_z + h_z \mu_0 \approx  - \frac{J_z+2h_z}{6}  \\
        & \times & \left(1 - \sqrt{1- \frac{12 J_a^2}{(J_z+2h_z)^2} } \right) \approx - \frac{J_a^2}{J_z+2h_z} \nonumber
\eea
consistent with what is expected from second order perturbation theory.

The energy eigenfunctions are given by $|n,p\rangle = \sum_{l=0}^\infty \psi_{n,l}(p)|p,l\rangle$ with coefficients
\be
\psi_{n,l} (p) \propto \left( \f{1-(-1)^l}{2} \right) J_{(l-\mu_n)/2}(x)
\ee
valid for $n$ odd, and
\bea
\psi_{n,l} (p) &\propto& \left( \f{1+(-1)^l}{2} \right) \left[ J_{(l-\mu_n)/2}(x)(1-\delta_{l,0}) \right. \nonumber \\
               &    &  \left. - \f{J_a}{E_n(p=0)} J_{(2-\mu_n)/2}(x) \delta_{p,0} \delta_{l,0} \right]
\eea
for $n$ even. Note that only the odd(even) $l$ coefficients are non-zero for odd(even) $n$.
For small $x$, the Bessel function is maximal when $l=\mu_n \approx n$. Qualitatively, this implies that for large magnetic fields, the $n$'th mode of the bound state is dominated by the state having $n$ overturned spins. This domination is total at $p=\pi/2,3\pi/2$ where $x=0$.

These energy wave functions are orthogonal when the variable $\mu_n$ obeys one of the
Eqs.~(\ref{eq_nu_odd}), (\ref{eq_nu_even}), or (\ref{pzeroroots}) due to the Bessel function property\cite{Prudnikov}
\begin{align}
  \sum_{k=1}^{\infty} & J_{k+\nu}(x) J_{k+\mu}(x) =  \\
  &= \frac{x}{2}\frac{J_{1+\nu}(x) J_{1+\mu}(x)}{\nu-\mu} \left(
\frac{J_{\nu}(x)}{J_{1+\nu}(x)} -
\frac{J_{\mu}(x)}{J_{1+\mu}(x)}\right). \nonumber
\end{align}

\section{Dynamic structure factor}
Having the eigenfunctions $|n,p\rangle$ and energies $E_{n}(p)$ we can calculate the dynamic structure factor, which at zero temperature is
\bea
    S^{\alpha \alpha'}(q,\omega) & = & \sum_{n} \langle 0,0| S^\alpha_{-q} | n,q \rangle \langle n,q | S^{\alpha'}_{q} | 0,0 \rangle  \nonumber \\
   &  & \times \delta(\omega - (E_n(q)-E_0(0)),
\eea
where $|0,0\rangle$ is the ground state and we have restricted the intermediate states to the states $|n,q\rangle$  in the zero and two domain-wall sectors that restricts the energy transfer $\omega < 2J_z$.  In the following we will consider $S^{xx}$, $S^{yy}$, and $S^{zz}$ separately.

For $S^{xx}$ and $S^{yy}$, it is convenient to consider the raising and lowering operators $S^\pm = S^x \pm iS^y$. Expressed in terms of these the transverse dynamic structure factors, $S^{xx}$ and $S^{yy}$ are
\begin{eqnarray}
  S^{xx} &=& \frac{1}{4}\left[(S^{+-}+S^{-+})+(S^{--}+S^{++})\right], \nonumber \\
  S^{yy} &=& \frac{1}{4}\left[(S^{+-}+S^{-+})-(S^{--}+S^{++})\right]. \nonumber
\end{eqnarray}
The ground state has zero momentum and can for $J_a\ll J_z$ be approximated by the ferromagnetic state $|FM \rangle$ where all spins point along the magnetic field.
The calculation is simplified greatly by this approximation as then the structure factors $S^{--}$, $S^{++}$, and $S^{+-}$ are zero, which follows from $S^+ |FM \rangle = 0$ implying that $S^{xx}=S^{yy}=\f{1}{4}S^{+-}$.  The action of $S^-$ on the ferromagnetic state creates a state with one down-spin, thus it belongs to the odd sector, and will have rather high energy, of the order $J_z+h_z$.
Using the eigenfunctions, we find
\begin{equation*}
    S^{+-}(q,\omega) = \sum_{n=1}^{\infty} \delta(\omega - E_{n}(q))\, I_n(q),
\end{equation*}
where $I_n$ is normalized relative intensity of the $n$'th mode,
\begin{equation} \label{Intensity}
    I_n(q) = \frac{|\psi_{n,l=1}(q)|^2} {\sum_l |\psi_{n,l}(q)|^2}.
\end{equation}
Using the expression for the wave functions and the following Bessel function identity
\begin{equation}
    \sum_{l=l_0}^{\infty} J^2_{l-\nu}(x) = - \frac{x}{2} J^2_{l_0-\nu}(x) \frac{\partial}{\partial \nu} \left[ \frac{J_{l_0-\nu-1}(x)}{J_{l_0-\nu}(x)} \right]
\end{equation}
with $l_0$ an integer, the intensity can be expressed in the form
\be \label{Intensity_formula}
  I_n(q) =   \left\{x \frac{\partial}{\partial \mu} \left[ \frac{J_{\mu/2}(x)}{J_{\mu/2+1}(x)} \right] \right\}^{-1} \bigg |_{\mu = -{\mu}_n},
\ee
where ${\mu}_n$ is the solution of Eq.~(\ref{eq_nu_odd}) for the odd domain length $l$.

For larger values of $J_a/J_z$, it is no longer adequate to approximate the ground state with the ferromagnetic state. Taking into account the exact nature of the ground state gives additional contributions to $S^{+-}$  and the corresponding intensity becomes
\begin{equation}
    I_n^{+-}(q) = \left( C^0_0 C_1^n(q)+ 2 \sum_{l>0} C^0_l C_{l+1}^n(q) \cos (ql/2) \right)^2, \label{I+-}
\end{equation}
where we have used the following notation for the normalized wave functions:
\begin{equation}
  C_l^n(q) = \frac{\psi_{n,l}(q)}{\sqrt{\sum_l |\psi_{n,l}(q)|^2}}.
\end{equation}
We will omit the momentum label for the ground state $C^0_l$ as it has zero momentum.
The leading terms of the intensity give the contribution
\begin{equation}
    I_n^{+-}(q) \approx (C^0_0)^2 I_n(q)+4 C^0_0 C^0_2 C_1^n(q)C_{3}^n(q) \cos q.
\end{equation}
In this case we also get non-zero contributions to $S^{-+}$ and to $S^{--},S^{++}$ which cause $S^{xx}$ to be different from $S^{yy}$. In the same notation as above their contributions are
\bea
    I_n^{-+}(q) & = & \left| C^0_1 C_0^n(0)\delta_{q,0}+ 2 e^{-iq/2}\sum_{l>1} C^0_l C_{l-1}^n(q) \cos (ql/2) \right|^2, \nonumber \\
   I_n^{++}(q) &=& \left( C^0_0 C_1^n(q) + 2 \sum_{l>0} C^0_l C_{l+1}^n(q) \cos(ql/2) \right) \nonumber \\
    & & \times \left( C^0_1 C_0^n(0)\delta_{q,0} + 2 \sum_{l>1} C^0_l C_{l-1}^n(q) \cos(ql/2) \right), \nonumber \\
  I_n^{--}(q) & = & (I_n^{++}(q))^*, \nonumber
\eea
where the $*$ means complex conjugation. Approximating these with their leading terms, we get
\bea
    I_n^{-+}(q) & \approx & \Big( C^0_1 C_0^n(0)\delta_{q,0}+2 C^0_2 C_1^n(q) \cos q \Big)^2,  \\
    I_n^{++}(q) &\approx& C^0_1 C_0^n(0) \Big( C^0_0 C_1^n(0)+2 C^0_1 C_2^n(0) \Big)\delta_{q,0} \nonumber \\
    & & + 2 C^0_0 C^0_2 (C_1^n(q))^2 \cos q.
\eea

Applying the operator $S^z$ to the ground state does not change the parity of $l$, thus all contributing intermediate states have even $n$. It is convenient to split off the ground state contribution as it has zero momentum and frequency, it represents the squared  magnetization, and write
\begin{eqnarray}
  S^{zz}(q,\omega) &=& \f{1}{4} \left[N - 2 \sum_l l (C^0_l)^2 \right]^2 \delta_{q,0} \delta(\omega) \\
     & & + \sum_n  \delta(\omega-E_n(q)) \, I_n^{zz}(q), \nonumber
\end{eqnarray}
where
\begin{equation} \label{I^zz_lead}
    I_n^{zz}(q) = \frac{ \left[ \sum_{l=2}^N C^0_l C_l^n(q) \sin(ql/2) \right]^2 }{\sin^2 (q/2)}.
\end{equation}
The leading contribution of the sum is
\begin{eqnarray}
  {I}_n^{zz} (q) &=& \frac{4 {E_0}^2}{J_a^2} \frac{|\psi_{n,l=2}(q)|^2}{\sum_l |\psi_{n,l}(q)|^2} \cos^2 (q/2) \nonumber \\
  & & \approx \frac{4 J_a^2}{(J_z+2h_z)^2} \cos^2 (q/2) \, {I}_n^{\mathrm{ev}}(q), \label{Izz_res}
\end{eqnarray}
where the introduced intensity ${I}_n^{\mathrm{ev}}$ corresponds to contribution from the states with even $n$ and can be written as
\begin{equation}
    {I}_n^{\mathrm{ev}}(q) = \left\{x \frac{\partial}{\partial \mu} \left[ \frac{J_{\mu/2}(x)}{J_{\mu/2+1}(x)} \right] \right\}^{-1} \bigg |_{\mu = -{\mu}_n},
\end{equation}
where ${\mu}_n$ is the solution of the Eq.~(\ref{eq_nu_even}) for the even domain length $l$.

\section{Finite temperature}
At finite temperatures, the dynamic structure factor will in addition to transitions out of the ground state also get contributions that depend solely on the excited states. In particular, there will be contributions at low frequencies corresponding to the spacing between energy levels. It was these temperature induced contributions that were the focus of the neutron scattering experiments in Refs.~\onlinecite{Christensen,Montfrooij}.  We will consider such temperature-induced contributions at relatively low frequencies $\omega < J_z$.

At finite temperature, entropic factors make it favorable to break up a spin domain, thus inducing domain-walls.  The magnetic field confines pairs of domain-walls leading to a picture of the finite temperature state as consisting of several bound states; short spin-down domains, interspaced by longer spin-up domains. To handle these thermal states, we will use the exact quantum mechanical treatment of an isolated bound state, and neglect the interaction between different bound states. We expect the quality of this noninteracting bound state approximation to be good on time scales shorter than the typical collision time between bound states. This collision time can be estimated by the mean distance between the bound states, which is the typical size of a spin-up domain, divided by the velocity of a bound state.
The typical length of a spin-up domain $\xi_{\up}$ in units of the lattice spacing can be estimated from the emptiness formation probability for the Ising model in a magnetic field\cite{Suzuki} and gives
\be
\xi_{\up}=\f{1}{1-\alpha_{\up}},
\ee
where
\be
\alpha_{\up} = \f{e^{\beta h_z/2}}{\cosh(\beta h_z/2) + \sqrt{ \sinh^2 (\beta h_z/2)+ e^{-\beta J_z}}},
\ee
where $\beta$ is the inverse temperature. Using this and the maximum velocity of a spin down bound state $v_{\rm max}$, we expect the independent bound state approximation to be good for frequencies
\be
     \omega > \f{2\pi v_{\rm max}}{\xi_{\up}}. \label{nonint_omega_criterion}
\ee
The bound state velocity $v_n(p) = \f{\partial E_n(p)}{\partial p}$ is largest for low-lying energy modes. For higher-energy modes, the dispersion becomes flatter and their velocity approaches zero. For the $n=1$ mode, the energy  varies as $E_1(p) \approx -2J_{\perp}[1+(J_a/J_\perp)^2] \cos p$ for $J_\perp > J_a$ at low momenta $p$, which implies a maximum velocity $v_1=2J_\perp[1+(J_a/J_\perp)^2]$. For $J_\perp < J_a$, the $n=1$ mode behaves almost as the $n=2$ mode, which for $2J_a/h_z < 1$ has a  maximum velocity $v_2 \approx 2J_a^2/h_z$ that increases for smaller fields and approaches $v_2 \sim 4J_a$.
Thus the maximum velocity of a bound state is $v_{\rm max} = {\rm max}(v_1,v_2)$.


The above validity criterion (\ref{nonint_omega_criterion}) takes into account the center of mass motion of the bound states. In addition, the quantum mechanical uncertainty in the size of a bound state can also ruin the noninteracting bound state approximation. From the wave functions, we estimate the size uncertainty to be $\pm 2J_a/h_z$, which implies that adjacent bound states have nonoverlapping boundaries when
\be
4J_a/h_z < \xi_{\up}. \label{nonint_boundaries_criterion}
\ee

Keeping in mind these restrictions, we can write down the dynamic structure factor at low frequencies $\omega < J_z$ in the independent bound state approximation as
\bea
  S^{\alpha \alpha'}(q,\omega) & = & \sum_{p,m,m^\prime} n_{m,p}  \mathbf{S}^{\alpha \alpha'}_{m^\prime m}(p,q)  \nonumber \\
 &  &  \times \delta \left[\omega-(E_{m^\prime}(p+q)-E_{m}(p)) \right],
\eea
where
\be
 \mathbf{S}^{\alpha \alpha'}_{m^\prime m}(p,q)
 =\langle m,p  | S^\alpha_{-q} | m^\prime,p+q \rangle \langle  m^\prime,p+q | S^{\alpha'}_{q} | m,p \rangle \nonumber
\ee
and $n_{m,p}$ is the occupation number of a bound state with internal energy index $m$ and momentum $p$. Its functional form depends generally on the statistics of these excitations, but is expected to behave at low temperatures as  $n_{m,p} \approx e^{-\beta (E_m(p)-E_0(0))} \kappa(\beta)$, where $\kappa$ is close to unity for $T<J_z$.

 The action of the operator $S^z$ on a state with a spin down cluster of $l$ spins and momentum $p$ is
\begin{equation}
    S^z_q |p,l\rangle = \f{N}{2} |p,l\rangle \delta_{q,0} - \frac{1-e^{i q l}}{1-e^{i q}} |p+q,l\rangle,
\end{equation}
which implies that the matrix elements $\mathbf{S}^{zz}_{mn}$ is given by the expression
\begin{eqnarray}
  \mathbf{S}^{zz}_{mn} &=& \delta_{q,0} \bigg( \f{N}{2}\sum_{l} C_l^n(p) C_l^m(p)-\sum_{l} l C_l^n(p) C_l^m(p) \bigg)^2  \nonumber \\
  &+& (1-\delta_{q,0})\frac{1}{\sin^2 (q/2)}\left( \sum_l C_l^n(p) C_l^m(p+q) \sin \frac{q l}{2} \right)^2.\nonumber \\
  &  & \label{Szzgeneral}
\end{eqnarray}
This expression is rather difficult to deal with analytically for general values of the parameters. However, in the region of parameters
where we expect the spectrum to be the WZL, we can evaluate it analytically.
Focusing on this region where $2J_a/h_z \ll 1$ and $J_\perp < J_a$, we get, see Appendix,
\begin{equation}\label{Szz_WZL}
    S^{zz}(q,\omega) =  \frac{\kappa(\beta) e^{-\beta (J_z+h_z)}}{1-e^{-\beta h_z}}\sum_{k=-N}^N G_k(q) \, \delta(\omega-2 h_z k),
\end{equation}
where the contribution from each mode for $q \neq 0$ is
\begin{eqnarray}
  G_0(q) &=& \frac{J_0^2(\zeta)}{\cosh(\beta h_z)-\cos q}\, \frac{e^{\beta h_z}+1}{2}, \nonumber \\
  G_k(q) &=& \frac{J_k^2(\zeta)}{2 \sin^2(q/2)}
    \begin{cases}
        1, \ \quad \qquad k>0,\\
        e^{\beta 2 h_z k},  \, \quad k<0,
    \end{cases} \nonumber
\end{eqnarray}
and the argument of the Bessel function is $\zeta=\frac{2 J_a}{h_z}|\sin q|$. For $q=0$, we get also a contribution from the ground state magnetization squared:
\begin{eqnarray*}
  G_0(0) &=& \left(\frac{N}{2}\right)^2-\frac{N}{1-e^{-\beta
h_z}}+\frac{1}{2}\,\frac{e^{\beta h_z}+1}{\cosh(\beta h_z)-1},  \\
  G_k(0) &=& \frac{1}{2} \left( \frac{2 J_a}{h_z}\right)^2
    \begin{cases}
        1, \ \quad \, \qquad k=1,\\
        e^{-\beta 2 h_z},  \, \quad k=-1, \\
        0, \ \quad \, \qquad |k|>1.
    \end{cases}
\end{eqnarray*}
Thus there are contributions for $\omega=2h_z k$, where $k$ is an integer. The $2h_z$ reflects the fact that $S^z$ does not change the parity of $l$. If we introduce the Bloch frequency $\omega_B = 2 h_z$, this gives a result similar to the expression found in Ref.~\onlinecite{KyriakidisLoss}. However, the temperature dependent factors are different. In particular, we get a pre-factor $e^{-\beta J_z}$, which is a consequence of the occupation number of bound states. This is in contrast to the factor $e^{-\beta J_z/2}$ expected in the single domain-wall approximation. A noteworthy feature of Eq.~(\ref{Szz_WZL}) is that increasing the magnetic field moves the spectral weight to lower Bloch frequencies. This follows from the fact that the maximum of the Bessel function squared $J_k(\zeta)$ for fixed $\zeta$ occurs when $k \approx \zeta-1$.

For the transverse dynamic structure factor, we find
\begin{eqnarray}
  \mathbf{S}^{+-}_{mn}(p,q) &=& \Bigg[ C_0^n(0) C_1^m(q) \delta_{p,0}  \label{S+-general}\\
  & & + 2 \sum_{l>0} C_l^n(p) C_{l+1}^m(p+q) \, \cos\left(\frac{q l -p}{2}\right) \Bigg]^2. \nonumber
\end{eqnarray}
This expression can also be evaluated analytically in the region where the spectrum is the WZL, see Appendix,
\begin{align}
    S^{+-}(q,\omega) &= \frac{\kappa(\beta) e^{-\beta J_z}}{e^{\beta h_z}-1} \sum_{k} \delta\big(\omega - h_z (2k+1)\big) \label{Spm_WZL}\\
    &\times 2 J_{k}^2 (\zeta)
    \begin{cases}
        1, \quad \qquad \qquad k>1/2,\\
        e^{\beta h_z (2k+1)},  \quad k<1/2,
    \end{cases} \nonumber
\end{align}
where $\zeta=\frac{2J_a}{h_z}|\sin q|$ and $k$ is an integer variable.
For the transverse structure factor, the excitations occur at frequencies that are an odd multiple of $h_z$, a consequence of the fact that $S^-$ changes the parity of $l$, the number of overturned spins.

\section{Cobalt chloride}
$\cobaltchloride$ is a quasi-one-dimensional anisotropic spin-1/2 magnet, proposed in Ref.~\onlinecite{KyriakidisLoss} as a candidate exhibiting BO in a magnetic field.  $\cobaltchloride$ has a dominant ferromagnetic coupling $J_z$ along the chains, which was determined from far-infrared absorption spectroscopy\cite{TorranceTinkham} to be $J_z = 36.5\mathrm{K}$. Other intrachain couplings $J_x$ and $J_y$ are smaller but non-zero.
The values of these couplings as well as other interchain couplings have been inferred both from far-infrared spectroscopy\cite{TorranceTinkham} and from spin wave analysis of neutron scattering experiments\cite{Kjems,Christensen,Montfrooij}. In this article, we use the following values to describe $\cobaltchloride$:
\begin{equation}\label{coupling_values}
    J_z = 36.5 \, \mathrm{K}, \ J_a = 3.8 \, \mathrm{K}, \ J_{\perp}= 5.43 \, \mathrm{K}.
\end{equation}
An important consequence of interchain couplings in $\cobaltchloride$  is that they cause the spins to order antiferromagnetically below $T_N=17.3 \, \mathrm{K}$. This implies that in the antiferromagnetic phase below $T_N$ the magnetic field $h_z$ used here should be interpreted as a sum of the external applied magnetic field and an internal field, which arises due to the magnetic moments of neighboring chains\cite{TorranceTinkham}.

\begin{figure}[tbp]
\begin{center}
    \includegraphics[width=8cm]{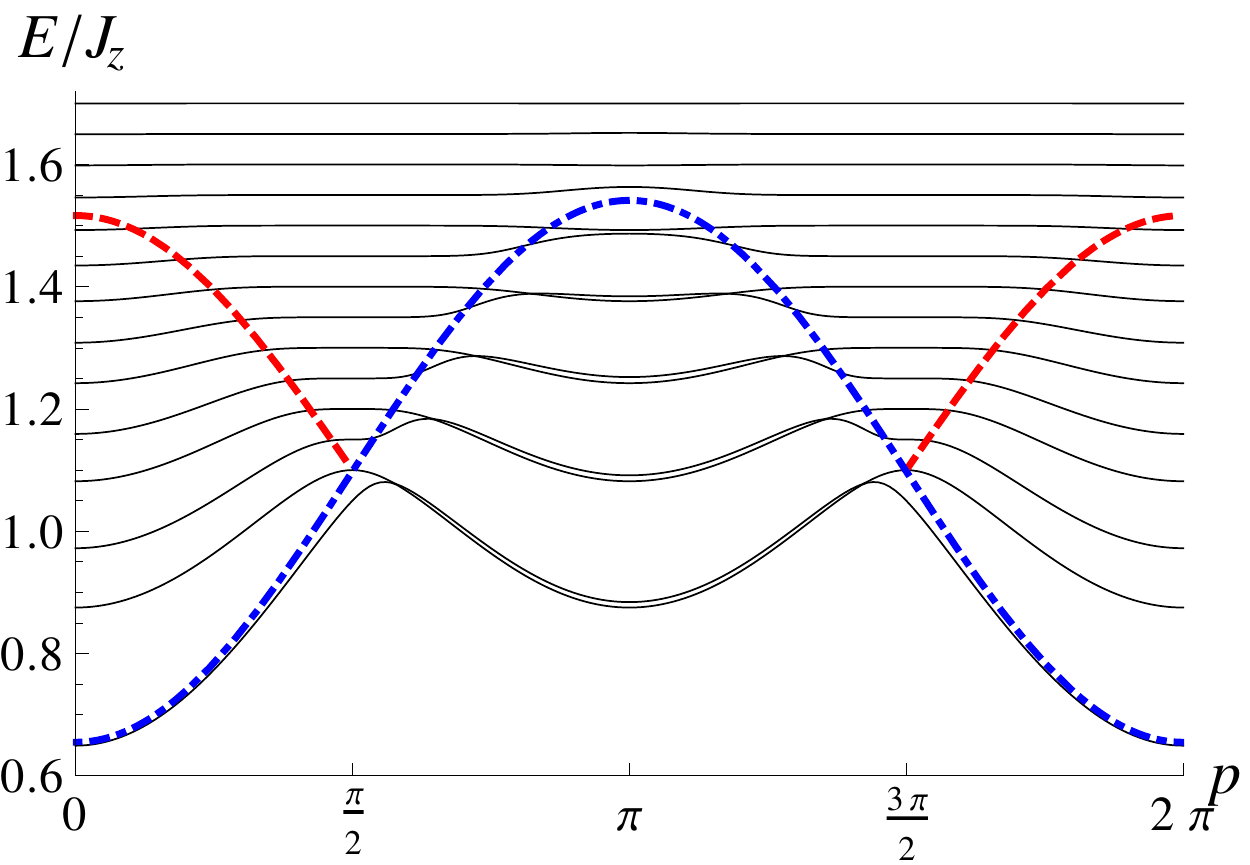} 
   \caption{(Color online) Energy levels vs. momentum computed with the parameters in Eq.~(\ref{coupling_values}) for a magnetic field $h_z/J_z = 0.05$.  The dot-dashed blue curve corresponds to the asymptotic dot-dashed line drawn in Fig.~\ref{rootsodd} but with $\gamma=1.43$, and the red dashed curve corresponds to the red dashed line in Fig.~\ref{rootsodd}.}
   \label{energylevelscobaltchloride}
  \end{center}
\end{figure}

We have plotted the energy levels $E_n(p)$ for the above couplings in Fig.~\ref{energylevelscobaltchloride}. The WZL is present at low energies in the momentum region around $p=\pi/2$ $(3\pi/2)$ and is bounded by the red dashed and blue dot-dashed curves, which correspond to the asymptotic lines drawn in Fig.~\ref{rootsodd}.
For energies $E > J_z+2h_z+4J_a$ the spectrum is the WZL for $p=0$ and for $E > J_z+h_z+2J_\perp[1+(J_a/J_\perp)^2]$, it extends also to the region above $\pi$ so that the spectrum is the WZL for all momenta. For regions of energies where the spectrum is not WZL for all momenta, it is possible to see from Fig.~\ref{energylevelscobaltchloride} that the even levels are symmetric around $\pi/2$ while the odd levels lack this symmetry property. This is a consequence of the skewness of levels seen in Fig.~\ref{rootsodd}.

\begin{figure}[tbp]
\begin{center}
    \includegraphics[width=8cm]{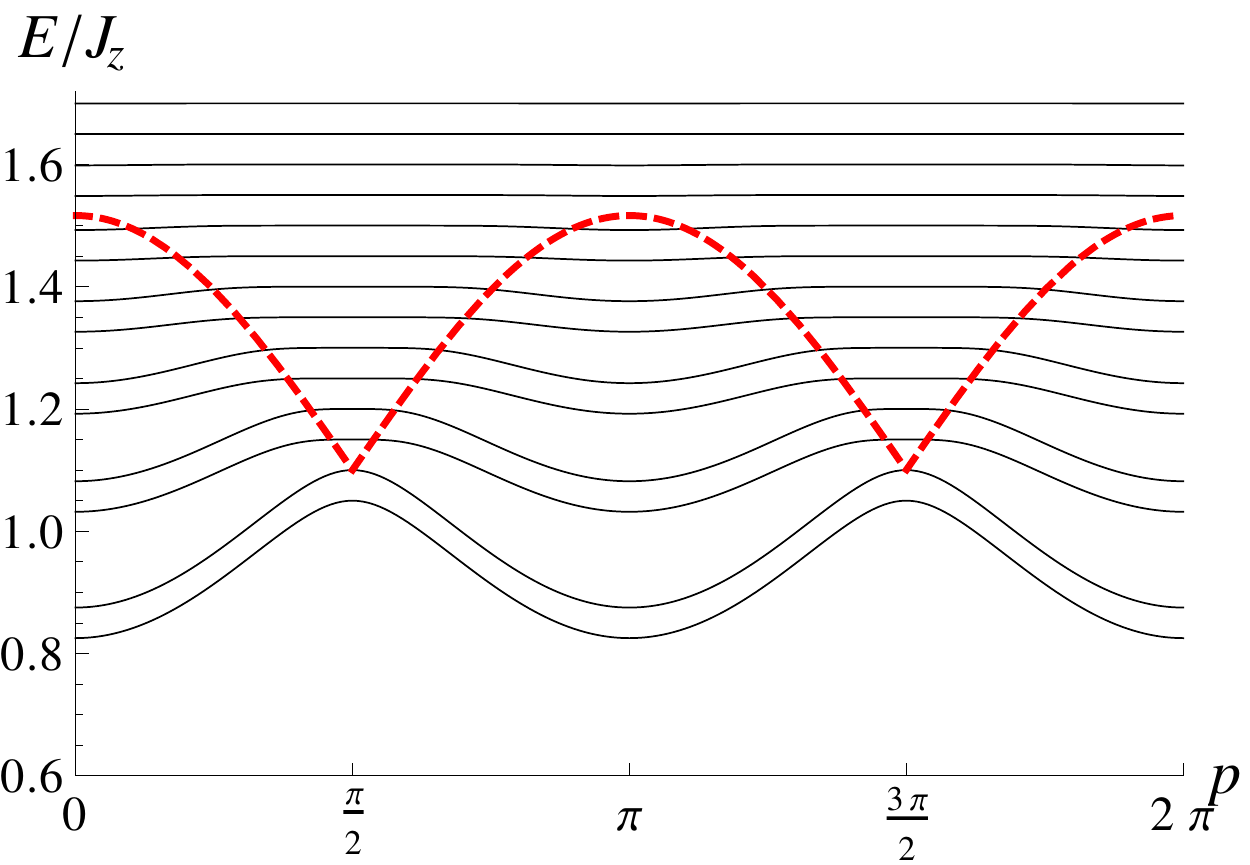} 
   \caption{(Color online) Energy levels vs. momentum computed with $J_\perp=0$, $J_a/J_z=0.104$, and $h_z/J_z = 0.05$.  The red dashed curve corresponds to the red dashed lines in Fig.~\ref{rootseven} and marks the lower boundary of the WZL.}
   \label{energylevelsJperpzero}
  \end{center}
\end{figure}

In order to see the effects of $J_\perp$, we have in Fig.~\ref{energylevelsJperpzero} also plotted the energy levels when $J_\perp=0$ for comparison\cite{Shiba}. We see that the main effect of $J_\perp$ is to lower the energy of the lowest odd level and to shift the low-energy odd levels in the region $\pi/2 < p < 3\pi/2$ so that they almost coincide with the even levels. The even levels are unaffected~by~$J_\perp$.

\begin{figure}[tbp]
\begin{center}
    \includegraphics[width=8cm]{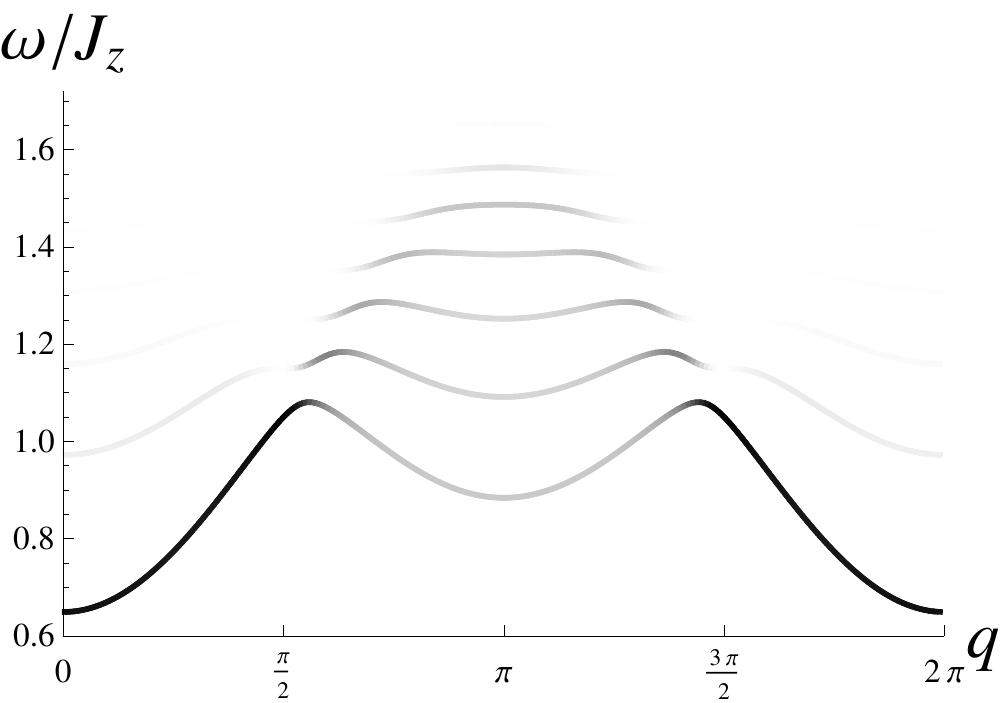} 
   \caption{Gray-scale plot of $S^{+-}(q,\omega)$ vs. $q$ and $\omega$ at $T=0$ for $J_\perp/J_z=0.149$, $J_a/J_z=0.104$, and $h_z/J_z=0.05$.}
   \label{SpmT0}
  \end{center}
\end{figure}

\begin{figure}[tbp]
  \begin{center}
    \includegraphics[width=8cm]{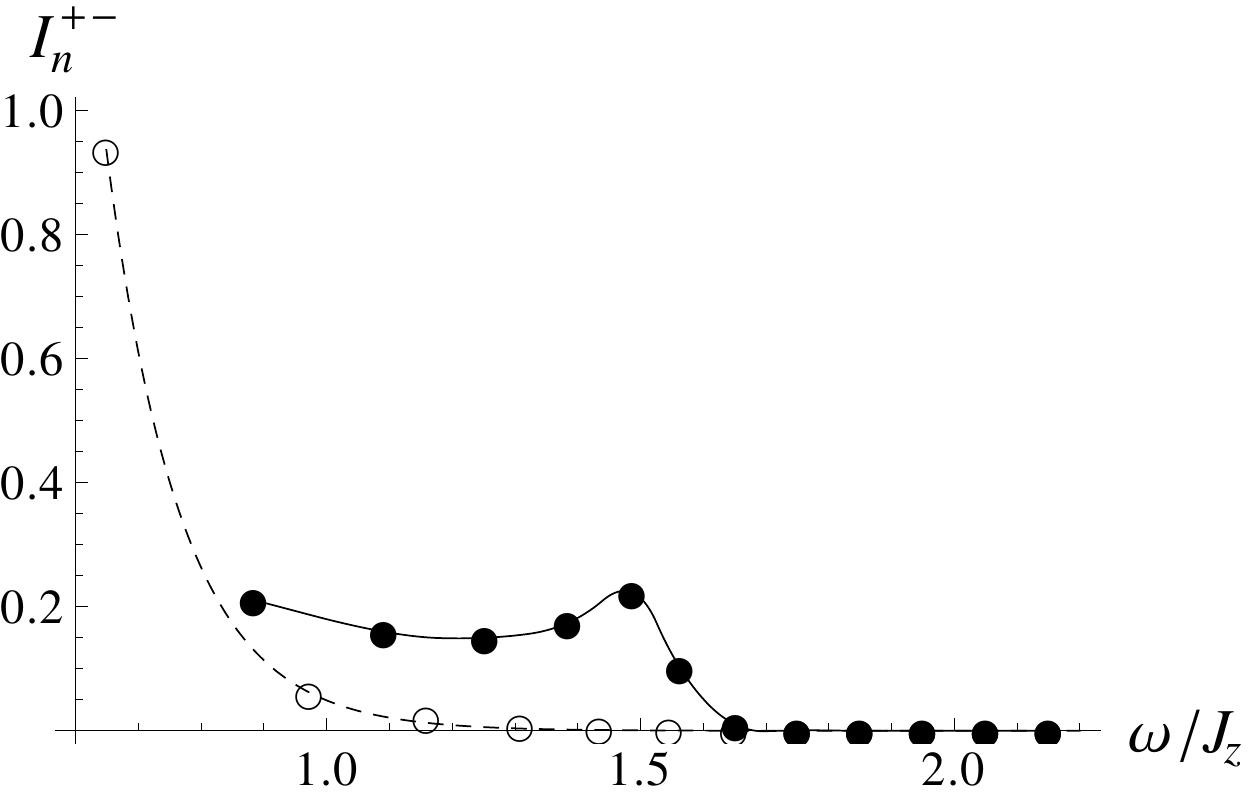} 
    \caption{Intensities $I^{+-}(q)$ vs. $\omega$ at $T=0$ for $J_\perp/J_z=0.149$, $J_a/J_z=0.104$, and $h_z/J_z=0.05$. The results for two momenta are shown, $q=0$ (open circles, dashed line) and $q=\pi$ (solid circles, solid line). The lines are guides to the eye.}
    \label{IntensityFig}
  \end{center}
\end{figure}

The zero temperature transverse dynamic structure factor $S^{+-}$ at $h_z/J_z=0.05$ is shown in
Fig.~\ref{SpmT0}. Only transitions to odd $n$ levels have nonzero intensity and it is seen that  most of the spectral weight occurs for transitions to the spin-wave like state $n=1$. The intensities of higher excited levels are weak in the momentum region around $p=\pi/2$ where we expect to see the WZL. Exactly at $p=\pi/2$, the size of the bound state is a good quantum number, thus higher excited bound states with $n>1$ have no amplitude to have the size $l=1$, which is the dominant intermediate state generated by neutron scattering on a ferromagnetic state. Any intensity of $n>1$ levels at $p=\pi/2$ reflects how the ground state deviates from being fully ferromagnetic. For the coupling constants relevant for $\cobaltchloride$, the probability for finding all spins up in the ground state is roughly $99\%$, so fluctuation corrections to the ground state are small and the integrated spectral intensity above the $n=1$ mode is less than $1\%$. In Fig.~\ref{IntensityFig}, we show how the intensities Eq.~(\ref{I+-}) of the different levels vary for two momenta $q=0$ and $q=\pi$. For $q=0$, the intensity drops exponentially with frequency, but for $q=\pi$, the intensity decreases only slightly before it increases up to the energy where the WZL sets in and then drops rapidly. This also reflects the fact that the main contribution comes from transitions to the $l=1$ state.

\begin{figure}[tbp]
\begin{center}
    \includegraphics[width=8cm]{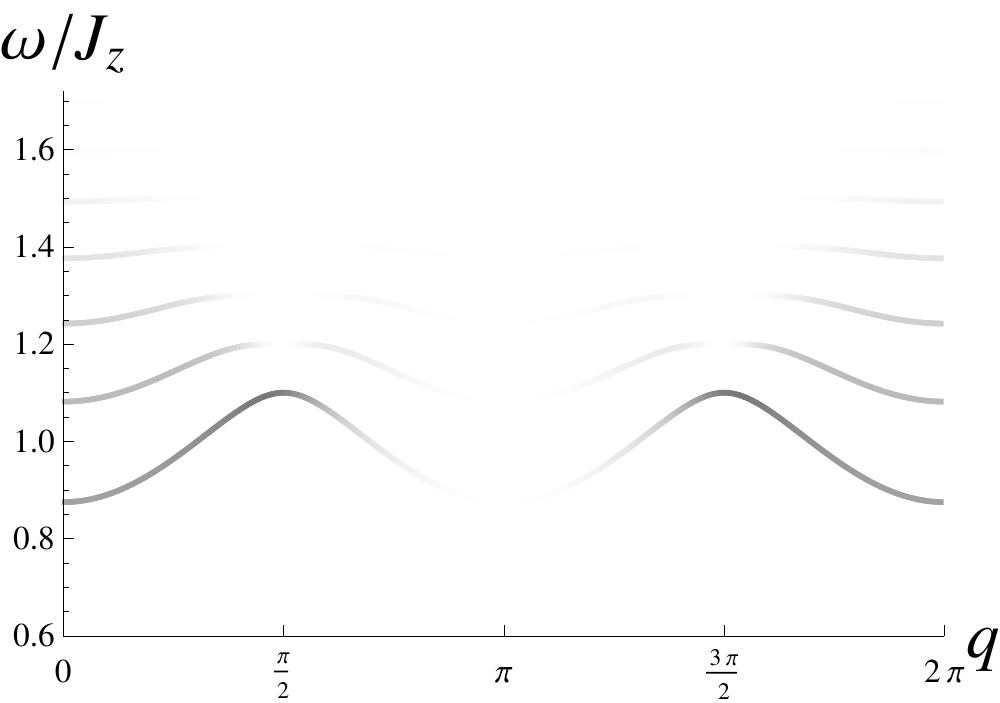} 
   \caption{Gray-scale plot of $S^{zz}(q,\omega)$ vs. $q$ and $\omega$ at $T=0$ for $J_\perp/J_z=0.149$, $J_a/J_z=0.104$, and $h_z/J_z=0.05$. The intensity of the plot has been increased by a factor $28$ in order to make it visible on the same gray-scale as used in Fig.~\ref{SpmT0}.}
   \label{SzzT0}
  \end{center}
\end{figure}

The behavior of the longitudinal dynamic structure factor $S^{zz}$ for parameters relevant for $\cobaltchloride$ is shown in Fig.~\ref{SzzT0}. Here, only excitations to even $n$ levels are nonzero which implies that $S^{zz}$ is independent of $J_\perp$. The total spectral weight of $S^{zz}(q\neq0)$ is however much smaller than for $S^{+-}$ because it is proportional to the probability for finding two overturned spins in the ground state. This is reflected by the small factor $J_a^2/(J_z+2h_z)^2$ in Eq.~(\ref{Izz_res}).

\begin{figure}[tbp]
\begin{center}
    \includegraphics[width=8cm]{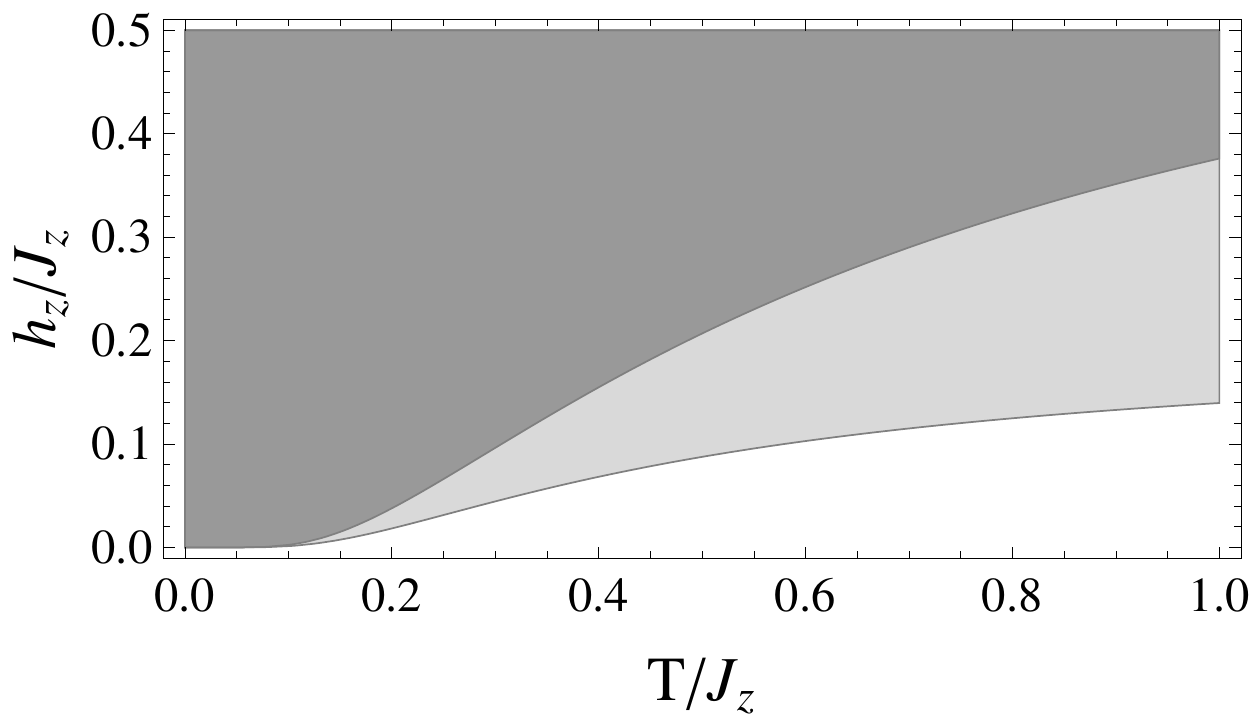} 
   \caption{Region of validity of the noninteracting bound state approximation using the parameters in Eq.~(\ref{coupling_values}). The region where the inequality (\ref{nonint_boundaries_criterion}) holds is shaded in light gray (which also overlaps entirely the dark gray region). The region where the noninteracting approximation can be used for frequencies down to the Bloch frequency $\omega_B=2h_z$, inequality (\ref{nonint_omega_criterion}), is shown in dark gray.}
   \label{validity}
  \end{center}
\end{figure}

For finite $T$, the validity of the noninteracting bound-state approximation for $\cobaltchloride$ used here is constrained mostly by $J_\perp$. Its relatively large value causes the $n=1$ bound state to have the largest velocity which according to the inequality (\ref{nonint_omega_criterion}) gives a lower bound on the frequency for which our approach is valid. If we require that this lowest frequency equals the Bloch frequency $\omega_B=2h_z$, the noninteracting bound-state approximation will be valid in the temperature/magnetic field region shaded dark gray in Fig.~\ref{validity} for the parameters relevant for $\cobaltchloride$. We do not expect our results to apply outside this region as a treatment of bound-state collisions is needed there.

In order to see finite temperature signatures of BO, a high temperature is needed to thermally occupy the bound state levels. However, at high temperatures, the validity of our approach is restricted to large magnetic fields, as seen from Fig.~\ref{validity}. Increasing the magnetic field has the disadvantage that the weights of the finite frequency Bloch peaks become small Eq.~(\ref{Szz_WZL}), thus weakening the signatures of BO. Therefore a judicial choice of temperature and magnetic field must be made to make observations possible.

\begin{figure}[tbp]
\begin{center}
    \includegraphics[width=8cm]{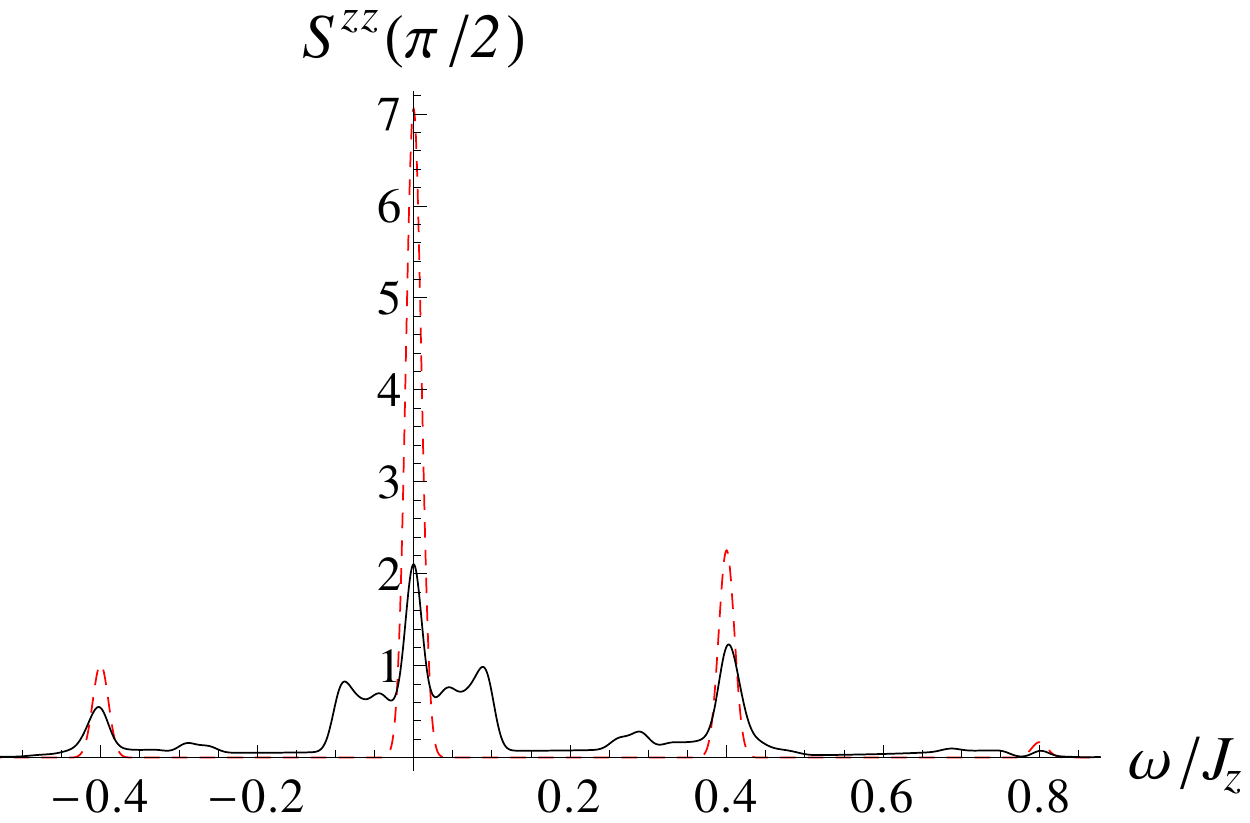} 
   \caption{(Color online) $S^{zz}(q=\pi/2,\omega)$ vs. $\omega/J_z$ for $T=J_z/2$, $h_z/J_z=0.2$, and parameters from Eq.~(\ref{coupling_values}). The red dashed curve is the corresponding WZL result Eq.~(\ref{Szz_WZL}) using the same parameters. In order to generate the plot delta-functions were approximated by a Gaussian distributions with variance $10^{-4}$. The vertical axis values are in multiples of $\kappa(\beta =2/J_z)$, a number of order unity.}
   \label{SzzT}
  \end{center}
\end{figure}

The optimal magnetic field for the first resonance at $\omega_B$ is $h_z \sim J_a$.
We will use a larger magnetic field, $h_z=0.2J_z$, as that allows our approach to be used up to a temperature $T \approx J_z/2$.
We find that for $S^{zz}$ the maximum intensity of the finite frequency WZL transitions occur at $q=\pi/2$. In Fig.~\ref{SzzT}, we have plotted our analytical result Eq.~(\ref{Szzgeneral}) numerically. We compare this with the expression obtained in the WZL limit Eq.~(\ref{Szz_WZL}) using the same parameters.
We see a clear peak at the Bloch frequency $\omega_B=2h_z$ also when the conditions for the WZL are suboptimal as is the case with the parameters in Eq.~(\ref{coupling_values}). The WZL calculation (red dashed line) overestimates the weight of the peaks, but do reasonably capture their relative intensities. For higher temperatures, the WZL expression matches Eq.~(\ref{Szzgeneral}) better as then more emphasis is put on the higher-energy part of the spectrum which is more WZL-like for all momenta. We wish to emphasize that the thermally induced transitions here come with an overall factor $e^{-\beta J_z}$ which makes them difficult to observe at low temperatures.

\begin{figure}[tbp]
\begin{center}
    \includegraphics[width=8cm]{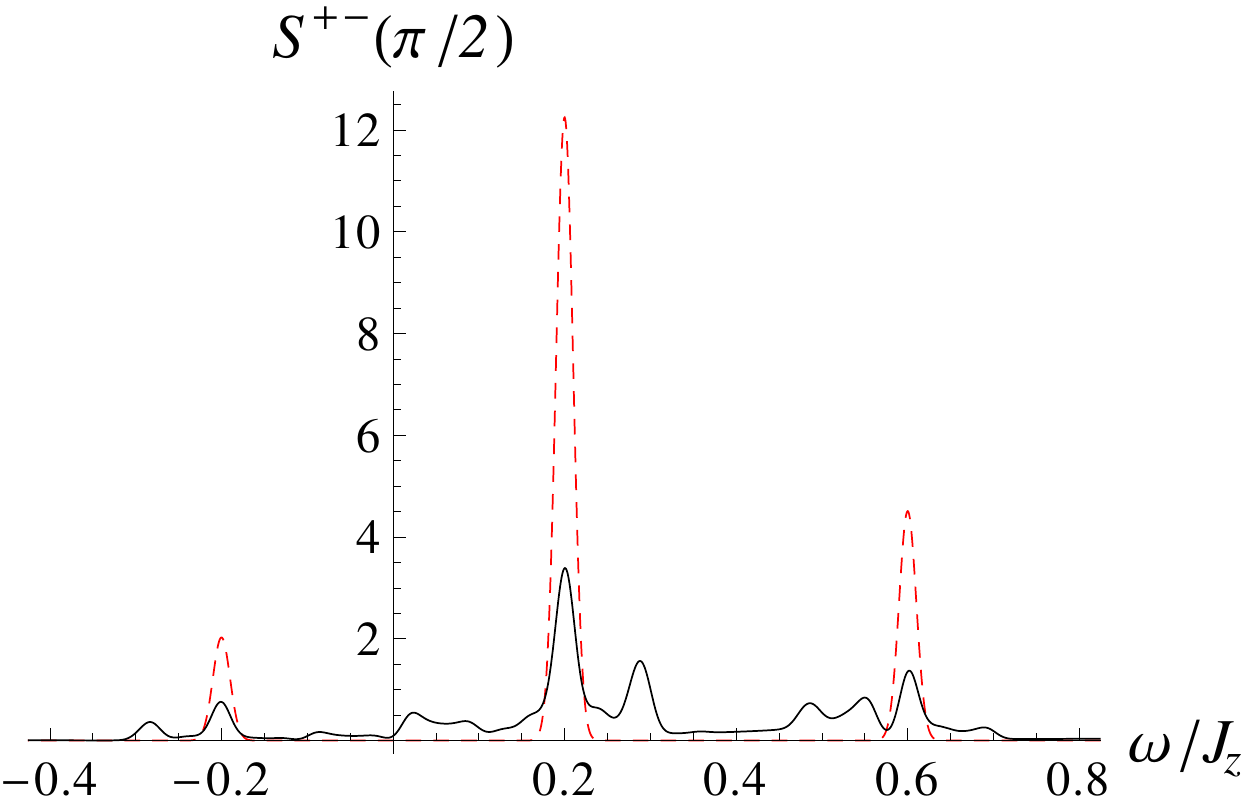} 
   \caption{(Color online) $S^{+-}(q=\pi/2,\omega)$ vs. $\omega/J_z$ using the same parameters as in Fig.~\ref{SzzT}. The red dashed curve is the corresponding WZL result Eq.~(\ref{Spm_WZL}) using the same parameters. Delta-functions were approximated by a Gaussian distributions with variance $10^{-4}$. The vertical axis values are in multiples of $\kappa(\beta =2/J_z)$, a number of order unity.}
   \label{SpmT}
  \end{center}
\end{figure}

In Fig.~\ref{SpmT}, we have plotted $S^{+-}$ using the same parameters as in Fig.~\ref{SzzT}.  Peaks at frequencies corresponding to odd multiples of the magnetic field are clearly seen among other peaks caused by the dispersion of the lowest-energy modes.

\section{Conclusion}
In this work, we have investigated the possibility of observing spectral signatures of magnetic BO in a one-dimensional anisotropic ferromagnetic spin system placed in a magnetic field. This system was considered previously\cite{KyriakidisLoss} but within an approximation where only a single domain-wall was included. We argue that the single domain-wall approximation is insufficient at a finite magnetic field. Instead, we consider a bound state of  a domain-wall and an anti-domain wall; a spin cluster of adjacent spins antialigned with the magnetic field\cite{TorranceTinkham}, and treat the thermal state as a noninteracting gas of such excitations. This allows us to also include the effects of the additional coupling $J_\perp$, which probably is present in most anisotropic materials that have a nonzero value of $J_a$. For instance in  $\cobaltchloride$ for which neutron scattering searches for BO have been made, $J_\perp$ is bigger than $J_a$.

We have treated the quantum mechanics of the bound state and obtained its energy levels and wave functions.
The spectrum in a magnetic field will be split by the magnetic field essentially into modes corresponding to the size (number of overturned spins) of the bound state. In the momentum region around $p=\pi/2$ ($p=3\pi/2$), the energy levels are equidistantly spaced down to the lowest energies. This corresponds to the WZL. For other momenta the exchange couplings, $J_a$ and $J_\perp$,  cause dispersion of the low-lying energy levels. This effect diminishes for higher energies and the spectrum becomes the WZL above a threshold energy for all momenta.

We have also calculated the neutron scattering dynamic structure factor at zero and low temperatures.
At zero temperature, the expected response occurs at high frequencies and there should be considerable chances of seeing the magnetic field splitting of the spectrum. However, it will be difficult to see the WZL because of the low spectral weight in this region. This is because neutron scattering flips a single spin and couples most strongly to the bound states that have a significant amplitude of having a single down-spin cluster, which has a dispersion that is heavily influenced by $J_\perp$. The longitudinal channel is not influenced by $J_\perp$, but is much weaker at non-zero momentum.

Detecting BO at finite temperatures with neutron scattering seems more promising. However, the thermal occupation number of bound states implies that these signatures will be suppressed at low temperatures as $e^{-\beta J_z}$. This might be the reason for the nonobservation of BO in the neutron scattering experiments\cite{Montfrooij,Christensen}. Nevertheless, it might still be possible that neutron scattering on $\cobaltchloride$ can be used to observe signatures of BO provided a careful selection of temperature and magnetic field is being made. In that respect, care must be taken so as to secure a big enough thermal population of bound states, a large intensity of the finite frequency resonance(s), and a regime where collisions of bound states do not destroy the BO.

\begin{appendix}
\section{WZL structure factors}
Let us calculate the dynamic structure factors at finite temperature in the limit where the spectrum is the WZL, that is $h_z \gg 2 J_a$ and $J_\perp \ll J_a$.
In this limit $E_n(p) = J_z + h_z n$, and the wave functions are Bessel functions of integer order
\begin{equation}\label{Cln_Jn}
    C_l^n(p) = \f{1+(-1)^{l-n}}{2} J_{\frac{l-n}{2}} ( x_0 \cos p ),
\end{equation}
where $x_0=2J_a/h_z \ll 1$. In the limit $x_0 \to 0$ these wave functions are normalized to unity. This also holds approximately at small finite $x_0$ which follows from the property of Bessel functions: $\sum_{k=-\infty}^{\infty} J_k^2(x) = 1$ and the fact that contributions from higher $J_k(x_0)$ decrease rapidly with $k$ for small $x_0$. Thus only small errors are introduced by extending the sum to $-\infty$ at small finite $x_0$.

For $S^{zz}(q\neq 0, \omega)$, the matrix elements (\ref{Szzgeneral}) become
\begin{align}\label{S^zz_mn}
    \mathbf{S}^{zz}_{mn} &\Big |_{q \neq 0} = \frac{1}{\sin^2 (q/2)} \\
    &\times \left[ \sum_{l>0} J_{\frac{l-n}{2}}( x_0 \cos p ) \, J_{\frac{l-m}{2}}\big(x_0 \cos (p+q)\big) \sin \left(\frac{q l}{2}\right) \right]^2, \nonumber
\end{align}
where the sum over $l$ goes over even(odd) integers when $n$ and $m$ both are even(odd), otherwise every term in the sum is zero. It is convenient to introduce a new integer-valued variable $t=(l-n)/2$ and rewrite the sum in the form
\begin{multline}
    \sum_{t>-\frac{n}{2}} J_{t}( x_0 \cos p ) \, J_{t+\frac{n-m}{2}}\big(x_0 \cos (p+q)\big) \sin \left[q\Big(t+\frac{n}{2}\Big)\right]\\
    =J_{\frac{n-m}{2}}(\zeta) \, \sin\left[\Big(p-\frac{\pi}{2}\Big)\frac{n-m}{2}+q\frac{n}{2}\right],
\end{multline}
where we introduced a new variable $\zeta = x_0 |\sin q|$. The sum over the product of Bessel functions was performed by extending the sum to negative $-\infty$ which only induces small errors when $x_0 \ll 1$, and then using Graf's addition theorem\cite{Abramowitz}
\begin{equation}
    \sum_{k=-\infty}^{\infty} J_{k+\nu}(u) \, J_k(v) \
    \begin{matrix}
        \sin\\
        \cos
    \end{matrix}
    (k \phi) = J_\nu(w) \
    \begin{matrix}
        \sin\\
        \cos
    \end{matrix}
    (\nu \chi)
\end{equation}
with the relations $w = \sqrt{u^2+v^2-2uv\cos\phi}$, $w \cos\chi = u-v\cos\phi$, and $w \sin\chi = v\sin\phi$.

The dynamic structure factor $S^{zz}$ for $q \neq 0$ in the independent bound state approximation can then be written in the form
\begin{align}\label{Szz_Loss_sum}
    S^{zz}(q,\omega)\Big |_{q \neq 0} &= \frac{\kappa(\beta)}{\sin^2 (q/2)}\sum_{m,n} e^{-\beta (J_z+h_z n)}\, \delta(\omega - (m-n)) \nonumber \\
    &\times J_{\frac{n-m}{2}}^2 (\zeta)
    \begin{cases}
        1/2, \ \qquad \quad n \neq m,\\
        \sin^2(q \frac{n}{2}),  \, \quad n=m,
    \end{cases}
\end{align}
where integration over momentum $p$ was performed for integer values of $n$ and $m$ variables
\begin{equation*}
    \int_0^{2\pi} \frac{dp}{2\pi} \, \sin^2\left[\Big(p-\frac{\pi}{2}\Big)\frac{n-m}{2}+q\frac{n}{2}\right] =
    \begin{cases}
        1/2, \ \qquad  n \neq m,\\
        \sin^2(q \frac{n}{2}),  \, n=m.
    \end{cases}
\end{equation*}
No transitions between the even and odd sectors are allowed, thus it is convenient to introduce a new integer-valued variable $k=(m-n)/2$ which describes the energy difference between the states involved in the transition.

We can reorder the double sum as
\begin{eqnarray}
    \sum_{m,n} &=& \big(\sum_{m \geq n}+\sum_{m<n}\big)\sum_n= \sum_{m \geq n}\sum_n  \\ \nonumber
    & & + \sum_m \sum_{n>m} = \sum_n \sum_{k \geq 0} +\sum_m \sum_{k < 0},
\end{eqnarray}
that allows us to rewrite the dynamic structure factor in the case of non-zero energy transitions, $k \neq 0$, in the following form
\begin{align}\label{Szz_Loss_k<>0}
    S^{zz}\Big |_{{q \neq 0} \atop {\omega \neq 0}} &= \frac{\kappa(\beta)}{2 \sin^2(q/2)} \, \frac{e^{-\beta J_z}}{e^{\beta h_z}-1} \\
    &\times \sum_{k \neq 0} \delta(\omega - 2 h_z k) J_{k}^2 (\zeta)
    \begin{cases}
        1, \ \qquad \quad k>0,\\
        e^{\beta 2 h_z k}, \quad \, k<0,
    \end{cases} \nonumber
\end{align}
where we used the expression for the sum of the first $N$ terms of a geometric series
\begin{equation}
    \sum_{n=1}^N e^{-\beta h_z n} = e^{-\beta h_z}\frac{1-e^{-\beta h_z N}}{1-e^{-\beta h_z}} \approx \frac{1}{e^{\beta h_z}-1}.
\end{equation}
For the zero mode, $k=0$, in order to find the sum in Eq.~(\ref{Szz_Loss_sum}) we can use the following identity
\begin{equation}
    \sum_{n=1}^{\infty} \sin^2(a n) e^{-b n} = \frac{1}{1-e^{-b}} \frac{\sin^2 a}{2} \frac{1+e^{-b}}{\cosh b-\cos 2a},
\end{equation}
which can be proved using Euler's formula and sum of terms of geometric series. This gives the contribution to the zero mode
\begin{equation}\label{Szz_Loss_k=0}
    S^{zz}\Big |_{{q \neq 0} \atop {\omega = 0}} = \kappa(\beta) e^{-\beta J_z} \frac{1+e^{-\beta h_z}}{1-e^{-\beta h_z}} \frac{\delta(\omega)}{2} \frac{J_0^2(\zeta)}{\cosh(\beta h_z)-\cos q}.
\end{equation}
Finally, combining together Eqs.~(\ref{Szz_Loss_k<>0}) and (\ref{Szz_Loss_k=0}) we obtain the dynamic structure factor:
\begin{equation}\label{Szz_Loss}
    S^{zz}(q,\omega)\Big |_{q \neq 0} = \frac{\kappa(\beta)e^{-\beta (J_z+h_z)}}{1-e^{-\beta h_z}}\sum_{k=-N}^N G_k(q) \, \delta(\omega-2 h_z k),
\end{equation}
where the contribution from each mode is
\begin{eqnarray}
  G_0 &=& \frac{J_0^2(\zeta)}{\cosh(\beta h_z)-\cos q}\, \frac{e^{\beta h_z}+1}{2}, \\
  G_k &=& \frac{J_k^2(\zeta)}{2 \sin^2(q/2)}
    \begin{cases}
        1, \ \quad \qquad k>0,\\
        e^{\beta 2 h_z k},  \, \quad k<0,
    \end{cases}
\end{eqnarray}
and the argument of the Bessel function is $\zeta=\frac{2 J_a}{h_z}|\sin q|$.

The leading contribution to the dynamic structure factor $S^{+-}$ comes from transition between states with nonzero momentum $p$. Then the matrix element in Eq.~(\ref{S+-general}) becomes
\begin{eqnarray}
    \mathbf{S}^{+-}_{mn}(p,q) &=& 4 \Bigg[\sum_{l>0} J_{\frac{l-n}{2}}(x_0 \cos p) \\
    & & \times J_{\frac{l+1-m}{2}}\big(x_0 \cos(p+q) \big) \, \cos\left(\frac{q l -p}{2}\right) \Bigg]^2, \nonumber
\end{eqnarray}
where the sum over $l$ is over even(odd) integers when $n$ is even(odd) and $m$ is odd(even).
Introducing the new integer variable $t =(l-n)/2$ and summing over the product of Bessel functions using Graf's addition theorem gives
\begin{multline}
    \sum_{t>-\frac{n}{2}} J_{t}( x_0 \cos p ) J_{t+\frac{n-m+1}{2}}\big(x_0 \cos (p+q)\big) \\
      \times \cos \left[q\Big(t+\frac{n}{2}\Big)-\frac{p}{2}\right]\\
    =J_{\frac{n-m+1}{2}}(\zeta) \, \cos\left[\Big(p-\frac{\pi}{2}\Big)\frac{n-m+1}{2}+\frac{p}{2}+q\frac{n}{2}\right].
\end{multline}
Since the only allowed transitions are between different parity sectors it is convenient to introduce the integer-valued quantity $k=(m-n-1)/2$. The integral over momentum is
\begin{equation*}
    \int_{0}^{2\pi} \frac{dp}{2\pi}\, \cos^2 \left[\Big( \frac{\pi}{2}-p \Big)k+\frac{p}{2}+q\frac{n}{2}\right] = \frac{1}{2}.
\end{equation*}
After reordering of the double sum we obtain finally
\begin{align}
    S^{+-}(q,\omega) &= \frac{\kappa(\beta)e^{-\beta J_z}}{e^{\beta h_z}-1} \sum_{k} \delta\big(\omega - h_z (2k+1)\big) \\
    &\times 2 J_{k}^2 (\zeta)
    \begin{cases}
        1, \quad \qquad \qquad k>1/2,\\
        e^{\beta h_z (2k+1)},  \quad k<1/2,
    \end{cases} \nonumber
\end{align}
where $\zeta=\frac{2J_a}{h_z}|\sin q|$ and $k$ is an integer.

\end{appendix}


\end{document}